\documentstyle[preprint,revtex,eqsecnum,cd]{aps}
\newcommand{\be}{\begin{equation}}   \newcommand{\ee}{\end{equation}}
\newcommand{\ba}{\begin{eqnarray}}   \newcommand{\ea}{\end{eqnarray}}
\newcommand{\bad}{\begin{eqnarray}\displaystyle}
\newcommand{\bA}{\[\begin{array}}    \newcommand{\eA}{\end{array}\]}
\newcommand{\bc}{\begin{center}}     \newcommand{\ec}{\end{center}}
\newcommand{\bb}{$\bullet\;$}        \newcommand{\eb}{$\;\bullet$}
\newcommand{\pr}{&\;\propto\;&}      \newcommand{\Prob}{{\cal P}(\Ef(E,x),E)}
\newcommand{\1}{\vspace{-1 mm}}      \newcommand{\2}{\vspace{-2 mm}}
\newcommand{\half}{\vspace{-.5 mm}}  \newcommand{\Dc}{\overline{D}}
\newcommand{\Ic}{\overline{I}}       \newcommand{\Jc}{\overline{J}}
\newcommand{\bra}{\langle}           \newcommand{\cket}{\rangle}
\newcommand{\Bra}{\left\langle\,}    \newcommand{\Cket}{\,\right\rangle}
\newcommand{\s}{\rm SPS}             \newcommand{\Dsf}{D^{(1)}_{\rm SPS}(E,x)}
\newcommand{\Dg}{\varrho_{\gamma}}   
\newcommand{\Dgf}{\Delta_1^{\gamma}} \newcommand{\Dgs}{\Delta_2^{\gamma}}
 \newcommand{\Dgn}{\Delta_n^{\gamma}}
\newcommand{\Dn}{\hat{\partial}_n}   \newcommand{\Df}{{\cal D}(E,x)}
\newcommand{\A}{{\cal A}(E)}         \newcommand{\T}{\varrho_{\gamma}x}
\newcommand{\Li}{\hat{\cal L}_i}     \newcommand{\Ec}{{\cal E}}
\newcommand{\Ei}{{\cal E}_i}         \newcommand{\Ef}{{\cal E}_1}
\newcommand{\K}{{\cal K}}            
\newcommand{\Y}{x''}                 \newcommand{\W}{E''}
\begin{document}
\begin{title}\\{\bf HIGH-ENERGY COSMIC-RAY MUONS UNDER\2\\THICK LAYERS OF
                  MATTER\1\\I. A Method to Solve the Transport Equation}\thanks
{To be published in the Proceedings of the Second NESTOR International
                    Workshop, 19 -- 21 October 1992, Pylos, Greece.}\end{title}
\author{{\bf V.~A.~Naumov}\2\thanks
                                 {Electronic mail address: naumov@fi.infn.it.}}
\begin{instit} Istituto Nazionale di Fisica Nucleare, Sezione di Firenze,
I -- 50125 Firenze\2\half, Italy \\ and Laboratory of Theoretical Physics,
               Irkutsk State University, 664003 Irkutsk,\2\1 Russia\end{instit}
\author{{\bf S.~I.~Sinegovsky}\2}
\begin{instit} Department of Theoretical Physics, Irkutsk State University,
                                         664003 Irkutsk,\2\1 Russia\end{instit}
\author{{\bf \'E.~V.~Bugaev}\2}
\begin{instit} Institute for Nuclear Research of the Russian Academy of
                     Science\2\half, \\ 117312 Moscow,\2\2\2 Russia\end{instit}

\begin{abstract} {\small An effective analytical method for calculating energy
\2 spectra of cosmic-ray muons at large depths of homogeneous media is
\2 developed. The method allows to include an arbitrary (decreasing) muon
\2 spectrum at the medium boundary and the energy dependence of both discrete
\2 (radiative and photonuclear) and continuous (ionization) muon energy losses,
\2 with resonable requirements for the high-energy behavior of the initial
   spectrum and differential cross sections of the muon-matter
   interactions\2\2\2\2\2\2.}\end{abstract}

 \narrowtext\newpage

\section{\bf Introduction}\label{s:I}

Cosmic-ray (CR) muons originate from the decay of unstable hadrons produced by
the interactions of cosmic-ray primaries and secondaries with nuclei of the
earth's atmosphere. Therefore the flux of CR muons contains information on
primary cosmic rays (energy spectrum, composition, anisotropy) as well as on
some properties of particle interactions at high and super-high energies.

During the last years the experimental investigations of CR muons with large
low-background detectors for penetrating particles have expanded rapidly in a
number of underground laboratories, in addition to the direct measurements in
the atmosphere and at ground level. Side by side with the traditional range of
problems of cosmic ray physics some additional aspects arise within the
framework of investigations with the new facilities. Thus, for example, the
flux of CR muons is used for calibration of the detectors and, at the same
time,
it is an important source of background events for the majority of underground
experiments, especially in neutrino astronomy and astrophysics. Detailed study
of this background is very important for further progress in astroparticle
physics.

Projects for the deep-underwater \v{C}erenkov and acoustic detection of
high-energy muons and neutrinos have been discussed for a long time. The
ultimate aim of these projects is to build detectors of volume $10^7\!-\!10^9$
m$^3$ or even larger \cite{BZ77,BBDGP90} which could be used, in particular, to
detect muons of energy up to $10^3$ TeV, in order to study the CR muon flux at
energies 2 to 3 orders of magnitude higher than those accessible in the present
experiments. It should be particularly emphasized that detectors of so enormous
volume can be used to accurately determine the energy, $E$, of individual muons
passing through the apparatus if $E$ exceeds a few TeV \cite{BZ77}.

In the near future the initial stage of the underwater muon and neutrino
telescopes DUMAND II in the ocean off the Hawaii island \cite{R92} and NT-200
in the Siberian lake Baikal \cite{Baikal92} will be commissioned. Besides,
several new programs have been proposed, such as a deep sea neutrino detector
NESTOR in the southwest corner of the Peloponnisos \cite{NESTOR92} and the
project AMANDA for a large scale muon and neutrino detector in deep ice at the
South Pole \cite{AMANDA91} (see also Ref.~\cite{S92} for a short summary on
next generation detectors). Precision calculations of different characteristics
of the CR muon flux after propagation through thick water layers are an
imperative element for the successful realization of these projects.

The transport of high-energy CR muons through dense media has been the subject
of theoretical investigations over many years with the use of analytical
\cite{ZM62,GZM76,N64,HNY64,N65,MN76,BNS84,BNS85IAN,Others}, numerical
\cite{NumMeth,LPKS91}, and Monte Carlo \cite{MonteCarlo,G84,TKAOM84} methods
(see also Refs.~\cite{MRM67,BKR70} for a review of the early literature). In
the majority of the papers listed the depth-intensity relation (DIR) was
studied. However, for future experiments with large-scale underwater neutrino
telescopes a detailed knowledge of the energy spectra of muons at very large
depths (large zenith angles) will be required in addition to the total
(oblique) intensities. Some results of calculations of the muon energy spectra
at large depths of matter were presented in Refs.~\cite{GZM76,G84,TKAOM84} and
in our previous papers \cite{BNS84,BNS85IAN}, but the increasing requirements
on accuracy of the calculations stimulate us to continue the investigation of
the problem.

The main difficulty in the calculation of muon intensity and spectrum at
large depths consists in the fact that an ultrarelativistic muon of energy $E$
above $\sim$ 100 GeV can, with comparable probabilities, lose in a single event
either a very small energy $\Delta E\ll E$ or an energy $\Delta E\sim E$ with
generation of a large electromagnetic or hadronic shower due to radiative or
photonuclear interaction with matter. These fluctuations of the energy loss
lead to a pronounced range straggling. Considering that the rate of radiative
and photonuclear losses increases with energy, the fluctuation effect grows
with energy and depth. An important consequence of this effect is the
impossibility to define a threshold energy for a muon reaching a given depth.
This fact presents a severe problem when reconstructing the surface (sea-level)
muon spectrum from a measured DIR (see, {\em e.g.},
Ref.~\cite{ACGK90})\footnote[1]
 {In this connection the problem of prompt muons which appear in the atmosphere
  due to decays of charmed hadrons should be mentioned
  \cite{PromptMuons,BNSZ89}. The data of the current underground experiments,
  those from European detectors (NUSEX, Frejus, MACRO), on the one hand, and
  those from Baksan and KGF, on the other hand, contradict each other (see
  Ref.~\cite{BNSZ89}). A certain part of these disagreements can be attributed
  to an inaccuracy in the computation of the fluctuation effect (see also
  Ref.~\cite{ACGK90} and a discussion in Ref.~\cite{KP91}).}.

Available exact analytical methods of cascade theory \cite{HNY64} require that
the initial muon spectrum at the boundary of the medium be a power-law and that
the differential cross sections for muon-matter interactions depend only on
the fraction of energy lost by the muon, $v = \Delta E/E$, but not on the muon
energy, $E$, itself (``scaling''). It is also assumed that the rate of the
continuous (ionization) energy loss is constant. We shall call this set of
assumptions the SPS model (Scaling + Power-law Spectrum). Such model have been
considered after Rozental' and Strel'tsov \cite{RS58} in
Refs.~\cite{ZM62,GZM76,N64}. Zatsepin and Mikhalchi \cite{ZM62} have suggested
a very simple approximate solution to the muon transport equation (TE). Their
approach has been generalized to a quasi-power initial spectrum \cite{GZM76}.
The exact solution for DIR within the framework of the SPS model has been
obtained by Nishimura \cite{N64} (see also Refs.~\cite{HNY64,N65,MN76}) with
the use of the technique of integral transformations. Both these approaches
have been employed with some modifications in numerous works (see, {\em e.g.},
Refs.~\cite{Others,TKAOM84}).

However, as is generally known, the assumptions of the SPS model hold roughly
only at very high muon energies (above $1\!-\!10$ TeV), and so the calculations
based on the SPS solution should be corrected by one or another way. Despite
the relatively weak energy dependence of the differential cross sections as
well as the closeness of the real sea-level muon spectrum to a power-law form
within wide energy intervals, these corrections prove to be very large and they
increase with depth. The point is that the muon energy spectrum under thick
layers of matter depends exponentially on integrals of the differential cross
sections with a weight which depends on the initial spectrum. To our knowledge
there is not any proper and consistent way to calculate these corrections at
large depths for the time being.

In the present study we discuss a comparatively simple and universal method for
the calculation of differential energy spectra as well as other important
characteristics of CR muons at arbitrary depth, which allows us to avoid from
the start the simplifying assumptions about the scale invariance of the cross
sections and the (quasi) power-law incident muon spectrum. The solution to the
TE is constructed by iterations, starting from an initial approximation with
the correct high-energy asymptotic behavior. In the range of applicability of
the initial approximation (sufficiently high energies) it becomes feasible to
introduce an (effective) analog of the threshold energy at the boundary which
is very useful in many respects. One of the advantages of the computer
realization of our approach (in comparison with the direct Monte Carlo
simulation or a purely numerical technique) is its high performance which
allows to carry out verifications of various hypotheses on the primary CR
spectrum and composition, charm production models, models of the photonuclear
interaction, {\em etc} with good precision and in a negligible CPU time. This
enables to estimate the sea-level muon spectrum using the data of
underground/underwater measurements by exhaustion, avoiding to solve the much
more difficult inverse scattering problem. 

It should be noted that the method under consideration is a development of our
previous studies \cite{BNS84,BNS85IAN}.

\newpage
The organization of this paper is as follows. In Sec.~\ref{s:P} we give some
preliminaries and notations. We present also a very short review on some
features of the differential cross sections for muon-matter interactions at
high energies which will be needed later on. In Sec.~\ref{s:CLA} the solution
to the TE in a continuous loss approximation is discussed briefly for the
methodological goals. In Sec.~\ref{s:SPS} we consider the exact solution to
the TE within the SPS model; for the present purpose (to study the asymptotic
behavior of the TE solution at high energies) the simplest expression in the
form of a series in powers of 1/E will be quite enough. In Sec.~\ref{s:FA} we
derive an approximate solution to the TE in the general case; essential
properties of the solution are discussed in some detail and illustrated by the
SPS model. The iteration algorithm for calculating corrections to the
approximate solution is described in Sec.~\ref{s:IS} and the convergency of the
algorithm is examined. Finally, in Sec.~\ref{s:SO} we summarize the results and
some perspectives for applications of the method.

\newpage\section{\bf PRELIMINARIES}\label{s:P}

\subsection{\bf Transport equation}\label{ss:TE}

The propagation of relativistic muons through a homogeneous medium is described
by the one-dimensional transport equation (TE)
\be\frac{\partial}{\partial x}D(E,x)-\frac{\partial}{\partial E}
             \left[\beta_i(E)D(E,x)\right] = \bra D(E,x)\cket\label{e:1.2.1}\ee
with the boundary condition \be D(E,0) = D_0(E)\;.\label{e:1.2.2}\ee
Here $D(E,x)$ is the differential energy spectrum of muons at depth $x$ in the
medium. In the general case \[x = \sec\vartheta\int_0^z\rho(z')dz'\;,\]
where $\rho(z)$ is the density of the medium at distance $z$ from the boundary,
and $\vartheta$ is the angle of incidence measured from the normal to the
boundary (zenith angle). The function $\beta_i(E) = -(dE/dx)_{\rm ion}$ is
the rate of the ionization energy losses which, as ever, are assumed to be
continuous. The symbol $\bra D\cket$ denotes a functional describing the
``discrete'' muon energy loss resulting from radiative and photonuclear
processes:  \be\bra D(E,x)\cket = \sum_k\bra D(E,x)\cket_k\;,\label{e:1.2.3}\ee
\bad\bra D(E,x)\cket_k & = & \Bra\frac{N_0}{A}
\int\frac{d\sigma_k^{Z,A}(E_1,E)}{dE}D(E_1,x)dE_1\Cket_{Z,A}\nonumber\\
&-&\Bra\frac{N_0}{A}\int\frac{d\sigma_k^{Z,A}(E,E_2)}{dE_2}D(E,x)dE_2
                                               \Cket_{Z,A}\;.\label{e:1.2.4}\ea
Here $d\sigma_k^{Z,A}(E_1,E_2)/dE_2$ is the differential cross section for a
muon interaction of type $k$: direct $e^+e^-$ pair production $(k = p)$,
bremsstrahlung $(k = b)$, and inelastic nuclear scattering $(k = n)$, and
$E_1$~($E_2$) is the initial (final) muon energy; $N_0$ is the Avogadro number;
$Z$ and $A$ are the atomic number and atomic weight of the target nucleus. The
brackets $\bra\ldots\cket_{Z,A}$ indicate an averaging over $Z$ and $A$.
Integrations in Eq.~(\ref{e:1.2.4}) are performed between the limits allowed by
the $k$-type process kinematics:
\[E^k_{1,\min}(E)\leq E_1\leq E^k_{1,\max}(E)\;,\;\;\;\;\;
                               E^k_{2,\min}(E)\leq E_2\leq E^k_{2,\max}(E)\;.\]

Equation (\ref{e:1.2.1}) does not take into account the muon finite lifetime,
which is permissible for ultrarelativistic energies and/or for dense enough
media\footnote[2]
 {The average decay range of a muon of energy $E$ is given by
  \[\lambda_d(E)=\tau_{\mu}p\rho/(m_{\mu}c)\simeq 6.23\times10^5\:{\rm g/cm^2}
    \left(\frac{\rho}{1\:{\rm g/cm^3}}\right)
    \left(\frac{p}{1\:{\rm GeV/c}}\right)\;,\] where $m_{\mu}$, $\tau_{\mu}$,
  and $p = \sqrt{(E/c)^2-(m_{\mu}c)^2}$ are the muon mass, lifetime, and
  momentum, respectively. Clearly $\lambda_d(E)$ is much longer than the muon
  ionization range $\lambda_i(E)$ \cite{PDG92} in a dense medium, so the muon
  decay effect is totally unessential in all instances of interest.}.
Moreover, in the equation (\ref{e:1.2.1}) (valid within the
``straight-forward'' scattering approximation) multiple Coulomb scattering and
the angular deflection due to inelastic scattering have been ignored. This
approximation is not so inoffensive but an examination of the problem does not
enter the scope of the present work.


A way to include the fluctuation effect due to knock-on electron production by
muons will be considered later on. An estimation of this effect for DIR has
been made by Nishimura \cite{N65}. According to Ref.~\cite{N65} the effect
leads to an increase of DIR at all depths by approximately 3\% (in the special
case of an initial spectrum $D_0(E)\,\propto\,E^{-4}$). A reliable analytical
method for describing the ionization straggling of relativistic muons with
incident energies below $\sim 100$ GeV in thick absorbers has been suggested
recently by Striganov \cite{St92}, but processes others than ionization were
not taken into account.

Let us introduce the macroscopic cross sections $\Sigma_k$ by the definition
\be\Sigma_k(v,E) = \Bra\frac{N_0}{A}\:
                \frac{d\sigma_k^{Z,A}(v,E)}{dv}\Cket_{Z,A}\;,\label{e:1.2.5}\ee
where
\[\frac{d\sigma_k^{Z,A}(v,E)}{dv} =
               \left|\frac{Ed\sigma_k^{Z,A}(E,E')}{dE'}\right|_{E'=(1-v)E}\;,\]
and $v$ is the fraction of energy lost. With help of Eq.~(\ref{e:1.2.5}) we
rewrite Eq.~(\ref{e:1.2.4}) in the more convenient form:
\be\bra D(E,x)\cket_k = \int^1_0[(1-v)^{-1}\Phi_k(v,E_v)D(E_v,x)-
                                      \Phi_k(v,E)D(E,x)]dv\;.\label{e:1.2.6}\ee
Here and below
\be\Phi_k(v,E) = \theta(v^k_{\max}(E)-v)
                      \theta(v-v^k_{\min}(E))\Sigma_k(v,E)\;,\label{e:1.2.7}\ee
$\theta(x)$ is the usual step function, $v^k_{\min}(E)$ and $v^k_{\max}(E)$
are the extreme values of $v$ for the $k$-type process, $E_v\equiv E/(1-v)$,
and the function $\Phi_k(v,E_v)$ is defined by Eq.~(\ref{e:1.2.7}) with the
substitution $E\Rightarrow E_v$.

At ultrarelativistic energies ($E\gg m_{\mu}c^2$, specifically at $E$ above
$\approx$ 10 GeV), we have with sufficient accuracy that
\be v^k_{\min}(E) = 0\;,\;\;\;\;v^k_{\max}(E) = 1\;,\label{e:1.2.8}\ee
and hence $\Phi_k(v,E) = \Sigma_k(v,E)$. Moreover, this may formally be
extended to all energies considering that radiative and photonuclear losses are
inessential to an accuracy of about 1\% at energies under 10 GeV for all media
of interest in cosmic ray physics \cite{PDG92,LKV85}, and only ionization
losses are important. Nevertheless, in the following we shall use approximation
(\ref{e:1.2.8}) only for asymptotic estimations, but we shall assume, if
necessary, that $v^k_{\min}(E)\ll 1$ and $1-v^k_{\max}(E)\ll 1$ for $k=p,b,n$
in the energy region covered.

\subsection{\bf Some features of muon-matter interactions
                                                 at high energies}\label{ss:CS}

A detailed description of the cross sections $d\sigma_k^{Z,A}(v,E)/dv$ used in
our calculations will be presented in a separate publication. For a short
review, see Ref.~\cite{LKV85}. To provide an inside into the properties of the
radiative processes we have presented in the Appendix a very simple
parameterization of the $v$-dependencies of the (normalized) cross sections
suggested by van Ginneken \cite{vG86}. As one can see from the Appendix,
strong energy loss fluctuations are more probable in bremsstrahlung. The direct
pair production cross section goes roughly as $1/v^2$ to $1/v^3$ over most of
the range ($v > 0.002$). Usually these losses are treated as continuous.
Nevertheless, as it follows from our estimations, the fluctuation effect
related to pair production is not exactly negligible and it can prove essential
at large depths. Hence we will be considering the pair production contribution
as discrete, together with bremsstrahlung.

To this must be added that in the limit of complete screening, i.e. for
\[\gamma_Z(v,E)\equiv\frac{200 q_{\min}}{m_eZ^{1/3}}\simeq\left(\frac{11}{Z}
           \right)^{1/3}\left(\frac{1\,{\rm TeV}}{E}\right)\frac{v}{1-v}\ll 1\]
(where $\gamma_Z$ is the degree of screening and $q_{\min}\simeq
m_{\mu}^2v/[2E(1-v)]$ is the minimum momentum transfer), the radiative cross
sections are functions of the variable $v$ only (scaling). However for values
of $v$ which are not too small (namely, at $1-v\ll 1$) complete screening
occurs only at very high energies, $E\sim 10$ TeV. At lower energies the cross
sections grow logarithmically with $E$.


Unfortunately there is no simple parameterization for the differential cross
section of the inelastic muon scattering on a nucleus $d\sigma_n^{Z,A}/dv$.
Moreover, both $v$- and (especially) $E$-behavior of the cross section are very
model dependent.

According to the vector-meson-dominance hypothesis $d\sigma_n^{Z,A}/dv$ is
expressed in terms of the total cross section for virtual photon absorption by
nucleons and nuclei. A generalized vector dominance model (GVDM) \cite{BB80}
adequately describes the features of these cross sections in the diffraction
region (low 4-momentum transfers, $Q^2$, and large photon energies, $\nu$):
growth with energy of the cross section for nucleon photoabsorption and
shadowing effects in nuclear photoabsorption. An approximate expression for
$d\sigma_n^{Z,A}/dv$ has been evaluated in the framework of the GVDM by
Bezrukov and Bugaev \cite{BB80}:
\[d\sigma_n^{Z,A}(v,E)/dv\propto\sigma_{\gamma N}(\nu)F_n(v,\nu)/v\;.\]
Here $\sigma_{\gamma N}(\nu)$ is the total cross section for absorption of a
real photon of energy $\nu = vE$ by a nucleon. In agreement with accelerator
and cosmic-ray experiments \cite{PhotCR1,PhotCR2} $\sigma_{\gamma N}(\nu)$
grows slowly above $\nu\sim 50$ GeV and can be represented approximately as
\[\sigma_{\gamma N}(\nu)\simeq\left[114.3 +
       1.647\ln^2\left(\frac{\nu}{47\,{\rm GeV}}\right)\right]\;\mu{\rm b}\;.\]
The growth of $\sigma_{\gamma N}(\nu)$ causes $d\sigma_n^{Z,A}/dv$ to depend on
the muon energy, $E$. The function $F_n(v,\nu)$ decreases slowly with
increasing $\nu$, gradually compensating the energy dependence of
$\sigma_{\gamma N}$ (a manifestation of the shadowing effect of nucleons
inside a target nucleus). Nevertheless, the logarithmic growth of
$d\sigma_n^{Z,A}/dv$ quantitatively remains up to $E\sim$ 10 TeV and possibly
in the asymptotics. 
The $v$-dependence of $F_n(v,\nu)$ is rather complicated; for $v$ over
$\sim 0.1$ it falls off roughly as $\ln v$ with increasing $v$, thus the
fluctuation effect due to this process is comparatively large.

It should be mentioned that the absence of any unitarity constraint allows a
very rapid (in comparison with the GVDM prediction) increase with energy of the
total photoproduction cross section as a result of the gluonic structure of the
high-energy photon (``minijet production mechanism'') \cite{Phot1}. Although
available cosmic-ray data obtained with underground detectors \cite{PhotCR1}
(for $\nu$ up to $\sim 10$ TeV) and with EAS arrays \cite{PhotCR2} ($\nu$ up to
$10^3\!-\!10^4$ TeV !) do not support this possibility, and what is more, a
recent study \cite{Phot2} has shown that the calculations of Ref.~\cite{Phot1}
strongly overestimate the minijet production contribution at $\nu > 10^3$ TeV,
a significant increase of the photoproduction cross section is still
anticipated at ultra-high energies. Thus the photonuclear interaction is one of
the interesting objects for study in future experiments with large underground
and underwater detectors.

\section{\bf CONTINUOUS LOSS APPROXIMATION}\label{s:CLA}

Let us at first consider the so-called continuous loss (CL) approximation
which is often-used for estimations of the CR muon intensity and spectrum under
thin enough layers of matter (see, {\em e.g.}, Refs.~\cite {BKR70,ACGK90} and
\cite{CLA}). It can be obtained from Eq.~(\ref{e:1.2.1}) by a formal expansion
of the integrand of expression (\ref{e:1.2.6}) in powers of $E_v$ at $E_v = E$,
to an accuracy of $O(v)$. As a result the functional (\ref{e:1.2.3}) becomes
\[\bra D(E,x)\cket =
        \sum_k\int^1_0(1+E\frac{\partial}{\partial E})\Phi_k(v,E)D(E,x)vdv\;.\]
Let us define
\[b_k(E) = \int^1_0\Phi_k(v,E)vdv =
                      \int^{v^k_{\max}(E)}_{v^k_{\min}(E)}\Sigma_k(v,E)vdv\;.\]
Clearly $b_k(E)$ is the relative partial rate of average energy loss due to the
$k$-type process, and
\[\beta(E) = \beta_i(E)+E\sum_k b_k(E) = -(dE/dx)_{\rm tot}\]
is the total rate of energy loss. Thus Eq.~(\ref{e:1.2.1}) in the CL
approximation takes the form
\be\frac{\partial}{\partial x}\Dc(E,x) =
             \frac{\partial}{\partial E}[\beta(E)\Dc(E,x)]\;.\label{e:1.3.1}\ee
Here and below $\Dc(E,x)$ stands for the differential muon spectrum in the CL
approximation. Similarly $\Ic(E,x)$ and $\Jc(x)$ will stand for integral
spectrum and DIR, respectively.

The solution to Eq.~(\ref{e:1.3.1}) with boundary condition (\ref{e:1.2.2}) is
\be\Dc(E,x)=D_0(\Ec(E,x))\frac{\beta(\Ec(E,x))}{\beta(E)}\;,\label{e:1.3.2}\ee
where $\Ec(E,x)$ is the (only) root of the equation $\lambda(\Ec,E) = x$, and
\[\lambda(E_1,E_2) = \int^{E_1}_{E_2}\frac{dE}{\beta(E)}\] is the average range
of a muon with initial energy $E_1$ and final energy $E_2$. In other words,
$\Ec(E,x)$ is the energy which a muon must have at the boundary of the medium
in order to reach depth $x$ with energy $E$. It is easily seen that $\Ec(E,x)$
is a monotonically increasing function of variables $E$ and $x$, and the
following identities are valid for any $x'\leq x$ and $E'\geq E$:
\[\Ec(\Ec(E,x'),x-x') = \Ec(E',x-\lambda(E',E)) = \Ec(E,x)\;.\]
It is clear also that $\Ec(E,0) = E$.

{}From Eq.~(\ref{e:1.3.2}) a very nice expression for the integral muon
spectrum
at depth $x$ can be obtained. Let $I_0(E)\equiv I(E,0)$ be the integral
spectrum at the boundary, then
\be \Ic(E,x) = \int^{\infty}_E\Dc(E',x)dE' =
        \int^{\infty}_{\Ec(E,x)}D_0(E')dE' = I_0(\Ec(E,x))\;.\label{e:1.3.3}\ee
According to Eq.~(\ref{e:1.3.3}) the expression for DIR, $J(x)$, can be
written as \be\Jc(x)=I_0(\Ec(E_t,x))\;,\label{e:1.3.4}\ee where $E_{t}$ is some
detection threshold. It can be argued that the value $\Jc(x)$ is practically
independent of $E_{t}$ at large depths when $E_{t}$ is sufficiently low
(really, when $E_{t}\ll 1$ TeV).

In spite of the simplicity and physical transparency of the CL approximation,
its range of application is fairly restricted. The inadequacy of this
approximation is obvious from the following simple example. Let the initial
spectrum, $D_0(E)$, have a breakoff at some energy $E_{\max}$, i.e. $D_0(E)=0$
at $E > E_{\max}$. Then, in accordance with Eqs.~(\ref{e:1.3.2}) and
(\ref{e:1.3.3}), $\Dc(E,x) = 0$ and $\Ic(E,x) = 0$ at $x >\lambda(E_{\max},E)$.
It is incorrect, of course, at least when $\lambda(E_{\max},E) < \lambda_d(E)$.
We will demonstrate below, within a simple model, that the CL solution has a
wrong asymptotic behavior as $E\rightarrow\infty$ and so it is irrelevant for
high energies.

\section{\bf ASYMPTOTIC BEHAVIOR (SPS MODEL)}\label{s:SPS}

We consider here the SPS model mentioned in Sec.~\ref{s:I}. Let us assume that
the functions $\Phi_k(v,E)$ and the ionization loss rate, $\beta_i(E)$, are
energy independent,
\be\Phi_k=\Phi_k(v)\;,\;\;\;\;\;\beta_i\equiv a = const\;,\label{e:i}\ee
and the initial spectrum is a power function of energy,
\be D_0(E) = D_0^{\gamma}(E) = CE^{-(\gamma+1)}\;.\label{e:ii}\ee
Moreover muon energies are assumed to be high enough so that conditions
(\ref{e:1.2.8}) is fulfilled.

In the SPS model the rate of the average energy loss is simply $a+bE$, where
$b = b_p + b_b + b_n$ is a constant\footnote[3]
  {In reality the quantities $b_k$ ($k = p, b, n$) and $\beta_i$ grow with
   energy logarithmically (or as a power of logarithm) up to $E\sim 10$
   TeV (see Refs.~\cite{PDG92,LKV85} and \cite{BB81}). But, as we have noted
   in Sec.~\ref{ss:CS}, it is not inconceivable, strictly speaking, that the
   growth of the relative rate of the average photonuclear loss extends at
   $E\gg 10$ TeV if a dramatic increase of the total photoproduction cross
   section with energy exists.},
and, therefore, the differential and integral muon spectra in the CL
approximation are described by
\be\Dc_{\s}(E,x) = D_0^{\gamma}(E)e^{-\gamma bx}\left[1+\frac{a}{bE}
                          (1-e^{-bx})\right]^{-(\gamma+1)}\;,\label{e:1.4.1}\ee
\be\Ic_{\s}(E,x) = I_0^{\gamma}(E)e^{-\gamma bx}\left[1+\frac{a}{bE}
                          (1-e^{-bx})\right]^{-\gamma}\;,    \label{e:1.4.2}\ee
where $I_0^{\gamma}(E)=\gamma^{-1}CE^{-\gamma}$ is the initial integral
spectrum.

A characteristic property of the SPS model in the CL approximation is a flat
(energy independent) spectrum (both differential and integral) for $E \ll W
\equiv a/b\sim 1$ TeV at sufficiently large depths $(x \gg 1/b)$,
\[\Dc_{\s}(E,x)\simeq D_0^{\gamma}(W)e^{-\gamma bx}\;,\;\;\;\;
  \Ic_{\s}(E,x)\simeq I_0^{\gamma}(W)e^{-\gamma bx}\;,\]
and recovery of its original form,
\[\Dc_{\s}(E,x)\simeq D_0^{\gamma}(E)e^{-\gamma bx}\;,\;\;\;\;
  \Ic_{\s}(E,x)\simeq I_0^{\gamma}(E)e^{-\gamma bx}\;,\]
for $E\gg W$ at all depths. According to Eq.~(\ref{e:1.3.4}), DIR takes the
form \[\Jc_{\s}(x) = I_0^{\gamma}(W(e^{bx}-1))\;,\] independently of the
threshold energy $E_{t}$ if $E_{t}\ll W(1-e^{-bx})$.

Let us consider now the exact solution to Eq.~(\ref{e:1.2.1}) within the
framework of the SPS model. Denote
\[b_{\gamma+n} = \int^1_0\Phi(v)[1-(1-v)^{\gamma+n}]dv\;,\;\;\;n=0,1,\dots\;,\]
and \[\Dg=b_{\gamma+1}-b_{\gamma}=\int^1_0\Phi(v)(1-v)^{\gamma}vdv\;,\] with
$\Phi(v) = \Phi_p(v)+\Phi_b(v)+\Phi_n(v)$. We shall seek the solution as a
series in powers of the dimensionless parameter $\xi = a/(\Dg E)$:
\be D_{\s}(E,x) = D_0^{\gamma}(E)e^{-b_{\gamma}x}\sum_{n=0}^{\infty}
                        \frac{(\gamma+1)_n}{n!}f_n(x)(-\xi)^n\label{e:1.4.3}\ee
(here $(\ldots)_n$ is the Pochhammer symbol). Substituting Eq.~(\ref{e:1.4.3})
into Eq.~(\ref{e:1.2.1}), we find that the coefficient functions $f_n(x)$
satisfy the following recurrence formula:
\be f'_n(x)+(b_{\gamma+n}-b_{\gamma})f_n(x) =
            n\Dg f_{n-1}(x)\;,\;\;\;\;\;f_n(0)=\delta_{n0}\;.\label{e:1.4.4}\ee
Integration of Eq.~(\ref{e:1.4.4}) yields
\be f_n(x) = \delta_{n0}+n\Dg\int^x_0\exp[-(b_{\gamma+n}
                         -b_{\gamma})(x-x')]f_{n-1}(x')dx'\;.\label{e:1.4.5}\ee
In particular, for $n = 0$ and 1 we have from Eq.~(\ref{e:1.4.5}) $f_0(x) = 1$
and $f_1(x) = 1-e^{-\T}$.

By induction, and using the fact that $b_{\gamma+n}-b_{\gamma} < n\Dg$ at
$n \geq 1$, one can easily verify that \[[f_1(x)]^n \leq f_n(x) \leq (\T)^n\]
for all values of $x$. Therefore the series (\ref{e:1.4.3}) is absolutely and
uniformly convergent under the condition
\be\zeta\equiv(\T)\xi = \frac{a x}{E}\leq 1\;,\label{e:1.4.6}\ee
but it is divergent when $\xi f_1(x) > 1$.

It can be shown that the obtained solution reduces to the solution in the CL
approximation (\ref{e:1.4.1}) if one sets formally $b_{\gamma+n}=(\gamma+n)b$,
for $n\geq 0$. A rough fulfillment of these equalities at not too large values
of $n$, which is a consequence of a quick growth of the electrodynamic cross
sections at $v\ll 1$ (see the Appendix), serves as the basis for the
applicability of the CL approximation.
It is obvious, however, that for any $n\geq 0$ and $\gamma > 1$ the exact
inequalities $b_{\gamma+n} < (\gamma+n)b$ take place, which are satisfied
(irrespective of the behavior of the function $\Phi(v)$), in so far as
$(1-v)^t > 1-tv$, at any $t > 1$ and $0 < v \leq 1$.

Thus the ratio \[r(E,x)=D_{\s}(E,x)/\Dc_{\s}(E,x)\;,\] which is a measure of
the fluctuation effect, increases with depth as $\exp[(\gamma b-b_{\gamma})x]$
at $\xi\ll 1$. In other words the CL approximation underestimates the muon
intensity at high energies. The magnitude of the effect depends critically on
the slope of the initial spectrum (the remainder $\gamma b-b_{\gamma}$ quickly
increases with $\gamma$) and it can be very large. To cite a single example,
$r(E,x)$ is about 10 at $E = 10$ TeV and $x = 10$ km of water equivalent (for
standard rock), in the case of the vertical spectrum of conventional CR muons
from the decay of $\pi$ and $K$ mesons \cite{BNS84}. It should be noted at the
same time that the ratio $r(E,x)$ does not necessarily exceed unit at all
energies.

The model under consideration shows that it is impossible to take into account
the fluctuation effect on muon spectra at large depths as a correction to the
CL approximation and a reliable method is required. Clearly the exact solution
(\ref{e:1.4.3}) by itself is unsuitable for calculations at fairly low energies
and/or at large depths; it cannot be used, in particular, to compute DIR. At
the same time the SPS model suggests a starting point for the required method:
we may, using an {\em ansatz} which has the correct asymptotic behavior at high
energies, construct the solution for the TE applying an appropriate iteration
procedure. In the next sections we will consider this approach to the problem.

It will be convenient to specify the asymptotic behavior of the cross sections
and initial spectrum at high energies as in the SPS model, that is to demand
the fulfilment of equalities (\ref{e:i}) and (\ref{e:ii}) at energies $E\gg
E_{\rm as}$, where $E_{\rm as}$ (a conventional bound of the asymptotic regime)
is a sufficiently large quantity. Then the SPS model will serve as a base for
asymptotic estimations. It will be recalled that the asymptotic form of the
photonuclear cross section contribution $\Phi_n(v,E)$ is actually unknown as
well as, strictly speaking, the high-energy behavior of the initial muon
spectrum, $D_0(E)$. Nevertheless, the condition imposed does not restrict
generality as long as a concrete value of the bound of the asymptotic regime,
$E_{\rm as}$, is not indicated. Evidently this condition does not play a part
in calculation of $D(E,x)$ at $E < E_{\rm as}$ due to the fast decrease with
energy of the initial spectrum.

\newpage\section{\bf GENERAL CASE: FIRST APPROXIMATION}\label{s:FA}

Consider the general case. Assuming analyticity of the ratio
$D(E_v,x)/D_0(E_v)$ as a function of the variable $v$ at the point $v = 0$, let
us expand this function in a power series in $v$. This yields
\[D(E_v,x)=D_0(E_v)\left[1+\sum^{\infty}_{n=1}v^n\Dn\right][D(E,x)/D_0(E)]\;,\]
where
\[\Dn \equiv \sum^n_{l=1}{{n\!-\!1}\choose{l\!-\!1}}
                           \frac{E^l}{l!}\:\frac{\partial^l}{\partial E^l}\;.\]
Then, introducing the definitions
\be\Delta_n(E) = \int^1_0\Phi(v,E_v)\eta(v,E)v^ndv\;,
                                   \;\;\;n = 1,2,\dots\,,\label{e:1.5.Delta}\ee
\be\A = \int^1_0 [\Phi(v,E)-\eta(v,E)\Phi(v,E_v)]dv\;,\label{e:1.5.A}\ee
with $\eta(v,E) = (1-v)^{-1}D_0(E_v)/D_0(E)$, we find
\be\bra D(E,x)\cket = \left[\sum^{\infty}_{n=1}\Delta_n(E)D_0(E)
                           \Dn D^{-1}_0(E)-\A\right]D(E,x)\;.\label{e:1.5.1}\ee

Due to the fact that the functions $\Phi_k(v,E)$ depend rather slowly
on $E$, and the initial muon spectrum $D_0(E)$ is close to a power-low one at
high enough energies (vide supra), the ratio $D(E,x)/D_0(E)$ should be
asymptotically a relatively slowly varying function of $E$. Thus the
derivatives $D_0(E)\Dn D^{-1}_0(E)D(E,x)$ are small. It is obvious also that
the integrals $\Delta_n(E)$ decrease with increasing $n$. Moreover, due to the
specific $v$-dependence of the cross sections (see Sec.~\ref{ss:CS}),
$\Delta_1(E)\gg\Delta_n(E)$ at $n > 1$. These simple heuristic considerations
allow us to use as a first approximation only two leading terms of the
expansion (\ref{e:1.5.1}). In this approximation Eq.~(\ref{e:1.2.1}) is merely
a partial differential one,
\be\left[\frac{\partial}{\partial x}-\beta_1(E)\frac{\partial}
             {\partial E}+{\cal R}(E)\right]D^{(1)}(E,x)=0\;,\label{e:1.5.2}\ee
where the following notations has been used:
\[\beta_1(E) = \beta_i(E)+\Delta_1(E)E\;,\;\;\;\;\;
                          {\cal R}(E) = \A-[g(E)+1]\Delta_1(E)-\beta'_i(E)\;,\]
with $g(E)+1 = -ED^{-1}_0(E)D'_0(E)$. We will assume subsequently that $g(E)$
is a positive definite and nondecreasing function. Clearly $g(E) = \gamma$ as
$E\gg E_{\rm as}$.

The solution to Eq.~(\ref{e:1.5.2}) can be expressed as
\be  D^{(1)}(E,x) = D_0(\Ef(E,x))\exp[-\K(E,x)] \equiv \Df\;,\label{e:1.5.3}\ee
where
\be\K(E,x) = \int^x_0{\cal R}(\Ef(E,x'))dx' = \int^{\Ef(E,x)}_E
                       \frac{{\cal R}(E')dE'}{\beta_1(E')}\;.\label{e:1.5.4}\ee

The function $\Ef(E,x)$ can be obtained from the equation $\lambda_1(\Ef,E)=x$
(an analog of the energy-range relation), with
\[\lambda_1(E_1,E_2) = \int^{E_1}_{E_2}\frac{dE}{\beta_1(E)}\;.\]
The properties of the function $\Ef(E,x)$ are completely similar to the ones of
above-mentioned function $\Ec(E,x)$, but the physical meaning of this quantity
is not so obvious. Considering that the function $\beta_1(E)$ is an effective
rate of the average energy loss (both continuous and discrete) for a given
initial muon spectrum, the function $\Ef(E,x)$ can be interpreted as the
effective (for the given $D_0(E)$) energy which a muon must have at the
boundary of the medium in order to reach depth $x$ having energy $E$
{\em with a nonzero probability}. To refine this interpretation let us rewrite
Eq.~(\ref{e:1.5.3}) in the form which is like the expression for the spectrum
in the CL approximation (\ref{e:1.3.2}):

\be\Df = D_0(\Ef(E,x))\frac{\beta_1(\Ef(E,x))}
                                        {\beta_1(E)}\Prob\;,\label{e:1.5.Df}\ee
where
\be{\cal P}(E_1,E_2)=\exp\left[-\int^{E_1}_{E_2}\frac{q(E')dE'}
                                         {\beta_1(E')}\right]\label{e:1.5.P}\ee
and
\be q(E) = {\cal R}(E)+\beta'_1(E) =
                         \A-g(E)\Delta_1(E)+\Delta'_1(E)E\;.\label{e:1.5.q1}\ee
Evidently the function $q(E)$ reflects the effect of muon range straggling. It
can be demonstrated that $q(E) > 0$ at least for high enough energies. Indeed,
substituting Eqs.~(\ref{e:1.5.Delta}) and (\ref{e:1.5.A}) into the right side
of Eq.~(\ref{e:1.5.q1}) yields
\ba                                                         q(E) &\;=\;&
\int^1_0\{\Phi(v,E)-[1+g(E)v]\eta(v,E)\Phi(v,E_v)\}dv\nonumber\\ &\;+\;&
\int^1_0[g(E_v)-g(E)]\Phi(v,E_v)\eta(v,E)vdv         \nonumber\\ &\;+\;&
\int^1_0 E_v\frac{\partial\Phi(v,E_v)}{\partial E_v}\eta(v,E)vdv\;.
                                                            \label{e:1.5.q2}\ea
The factor $[1+g(E)v]\eta(v,E)$ does not exceed unit\footnote[4]
 {Because the derivative \[\frac{\partial}{\partial v}\{[1+g(E)v]\eta(v,E)\} =
   -g(E_v)\left\{1-\frac{g(E)}{g(E_v)}+[1+g(E)]\frac{v}{1-v}\right\}\eta(v,E)\]
 is negative for any $v > 0$ and $\eta(0,E) = 1$.}
and decreases fast (tends to zero) with increasing $v$, while the function
$\Phi(v,E_v)$ depends on the second argument, $E_v$, only logarithmically.
Thus the first integral in Eq.~(\ref{e:1.5.q2}) is positive. The second
integral is nonnegative on the assumption that $g(E)$ is an increasing (or
constant) function. The third integral is small in comparison with the first
one due to the factor $\eta(v,E)v$ in the integrand and (mainly) to the
inequality
\[E_v\left|\frac{\partial\Phi(v,E_v)}{\partial E_v}\right|\ll\Phi(v,E_v)\;,\]
which takes place even in the absence of the full screening [notice that
$\gamma_Z(v,E_v) < 1$ at $E$ above $\sim$1 TeV at any $v$]. Hence the last
contribution cannot change the sign of the function $q(E)$.

Thus the function $q(E)$ can be interpreted as an effective absorption
coefficient dependent upon the radiative and photonuclear energy losses, and
the function $\Prob$ should be treated as the probability for a muon with
energy $\Ef(E,x)$ at the surface to reach depth $x$ with energy $E$.

Simple examination shows that $\Delta_1(E) < b(E)$. Therefore,
$\Ef(E,x) < \Ec(E,x)$ for all values of $E$ and $x$. Moreover, the remainder
$\Ec(E,x)-\Ef(E,x)$ increases fast with depth since
\[\frac{\partial}{\partial x}[\Ec(E,x)-\Ef(E,x)]
                                          = \beta(\Ec(E,x))-\beta_1(\Ef(E,x))\]
\[              = \beta_i(\Ec)-\beta_i(\Ef)+b(\Ec)\Ec-\Delta_1(\Ef)\Ef > 0\;,\]
and we have taken into account that $\beta_i(E)$ is a nondecreasing function
of $E$ after a broad minimum at $p\approx 300$ MeV/c, almost independently of
the medium \cite{PDG92}. It is obvious also that the remainder $\Ec(E,x)-
\Ef(E,x)$ increases when the slope of the initial muon spectrum grows. The
decrease of the minimal muon energy at the surface, necessary in order that a
muon can reach a given depth with a given energy, is an evident reflection of
the discreteness of radiative and photonuclear muon energy losses. The function
$\Ef(E,x)$ is a useful approximation to estimate this minimal energy within the
scope of the approximate solution (\ref{e:1.5.3}).

{}From Eqs.~(\ref{e:1.5.Df}-\ref{e:1.5.q1}) the following expression for the
integral spectrum in the first approximation can be obtained:
\[I^{(1)}(E,x) = \int^{\infty}_{\Ef(E,x)}D_0(E'){\cal P}(E',\Ef(E',-x))dE'\;,\]
which is an evident generalization of Eq.~(\ref{e:1.3.3}) obtained in the CL
approximation. It is obvious that $I^{(1)}(E,x) < I_0(\Ef(E,x))$. In the
realistic case, when $q(E)$ is a function slowly varying with energy we find
\[I^{(1)}(E,x)\simeq I_0(\Ef(E,x))e^{-\overline{q}x}\;,\]
where $\overline{q}$ is the average of $q(E)$.

Consider now the approximate solution (\ref{e:1.5.3}) in the SPS model. It is
clear that all moments
\[\Delta_n = \int^1_0\Phi(v)(1-v)^{\gamma}v^ndv\equiv\Dgn\]
(in particular, $\Delta_1 = \Dgf\equiv\Dg$) and the parameter ${\cal A} =
b_{\gamma}$ are constant in this case. So the effective absorption coefficient
$q = b_{\gamma}-\gamma\Dg\equiv q_{\gamma}$ is a positive constant, such that
\[\frac{1}{2}\gamma(\gamma+1)\Delta^{\gamma}_2 < q_{\gamma}
                                                      < \gamma^2\Delta^0_2\;.\]
One can easily show that
\[\Ef^{\s}(E,x) = E[(1+\xi)e^{\T}-\xi]\;\;\;{\rm and}\;\;\;
                             {\cal P}(\Ef^{\s}(E,x),E) = e^{-q_{\gamma} x}\;.\]
By this means the differential and the integral spectra can be written as
\be\Dsf = D_0^{\gamma}(E)e^{-b_{\gamma}x}
               \left[1+\xi(1-e^{-\T})\right]^{-(\gamma+1)}\;,\label{e:1.5.5}\ee
\be I_{\s}^{(1)}(E,x) = I_0^{\gamma}(E)e^{-b_{\gamma}x}
                   \left[1+\xi(1-e^{-\T})\right]^{-\gamma}\;,\label{e:1.5.6}\ee
and DIR becomes
\[J_{\s}^{(1)}(x) = I_0^{\gamma}(W_{\gamma}(e^{\T}-1))e^{-q_{\gamma}x}\;,\]
for any $E_{t}\ll W_{\gamma}(1-e^{-\T})$, where $W_{\gamma} = a/\Dg$. Thus,
at large depths,
\[J_{\s}^{(1)}(x)/\Jc_{\s}(x)\simeq(\Dg/b)^{\gamma}e^{(\gamma b-b_{\gamma})x}\]

As might be expected, the first two terms of the exact $1/E\,$-expansion
(\ref{e:1.4.3}) coincide with the corresponding terms of the same expansion for
$\Dsf$. Therefore the approximation (\ref{e:1.5.5}) has the correct behavior at
high energies at least when $\zeta\leq 1$ (see (\ref{e:1.4.6})). It should be
noted also that the approximation (\ref{e:1.5.5}) is self-consistent. Indeed,
one can show that
\[\Dn\left[\frac{\Dsf}{D_0^{\gamma}(E)}\right] = \frac{(\gamma+1)_n}
                     {n!}Z^n(E,x)\left[\frac{\Dsf}{D_0^{\gamma}(E)}\right]\;,\]
where
\[Z(E,x) = \frac{\xi f_1(x)}{1+\xi f_1(x)}
                                 = \frac{\xi(1-e^{-\T})}{1+\xi(1-e^{-\T})}\;.\]
Considering that $Z(E,x) < 1$ at any $E$ and $x$ the series in the right side
of Eq~(\ref{e:1.5.1}) is always uniformly convergent and one may actually cut
it off after the 1st term if
\[\frac{(\gamma+2)}{2}\,\frac{\Dgs}{\Dgf}\,Z(E,x) \ll 1\;,\]
and that is certainly admissible when $\xi f_1(x) \ll 1$. This is supporting
the approximation (\ref{e:1.5.3}) as a suitable ansatz.

\section{\bf GENERAL CASE: ITERATION SCHEME}\label{s:IS}

Let us now represent the solution to Eq.~(\ref{e:1.2.1}) by the following form
\be D(E,x) = D^{(1)}(E,x)[1+\delta(E,x)]\;,\label{e:1.6.1}\ee
where $\delta(E,x)$ is an unknown function (``relative correction''). To
derive the equation for $\delta(E,x)$ it is convenient at first to rewrite
Eq.~(\ref{e:1.5.2}) as
\be\frac{\partial}{\partial x}\Df-\frac{\partial}{\partial E}
            [\beta_i(E)\Df]=[\Delta_1(E)\omega(E,x)-\A]\Df\;,\label{e:1.6.2}\ee
where
\be\omega(E,x) = \frac{E[h(E)-h(\Ef(E,x))]}{\beta_1(E)}\;,\label{e:1.6.3}\ee
 with
\be h(E) = {\cal R}(E)+\frac{[g(E)+1]\beta_1(E)}{E} =
               \A+\frac{[g(E)+1]\beta_i(E)}{E}-\beta'_i(E)\;.\label{e:1.6.4}\ee
In order to derive Eq.~(\ref{e:1.6.2}) we have used Eqs.~(\ref{e:1.5.3}) and
(\ref{e:1.5.4}). Direct substitution of Eq.~(\ref{e:1.6.1}) into
Eq.~(\ref{e:1.2.1}), in view of Eq.~(\ref{e:1.6.2}), then gives
\ba\Li\delta(E,x)&=&\int^1_0\Phi(v,E_v)\left\{\Omega(E,x;v)
         [1+\delta(E_v,x)]\right.\nonumber\\&-&\left.[1+\omega(E,x)v]
                        [1+\delta(E,x)]\right\}\eta(v,E)dv\;,\label{e:1.6.5}\ea
where the differential operator
\[\Li = \frac{\partial}{\partial x}-\beta_i(E)\frac{\partial}{\partial E}\]
was introduced, and we have defined
\be\Omega(E,x;v) = \frac{D_0(E)}{D_0(E_v)}\,\frac{D_0(\Ef(E_v,x))}
                  {D_0(\Ef(E,x))}\,\exp[\K(E,x)-\K(E_v,x)]\;.\label{e:1.6.6}\ee
Clearly $\delta(E,0) = 0$. We shall seek the solution to Eq.~(\ref{e:1.6.5})
using a procedure of successive approximations.

Let us note initially that the function $\delta(E,x)$ follows a
$c_2(x)/E^2\,$-dependence as $E\gg E_{\rm as}$, where $c_2(x)$ is independent
of energy. It is a straight corollary of the coincidence of the first two terms
in the $1/E\,$-expansions for the approximate solution (\ref{e:1.5.3}) and the
exact SPS solution (\ref{e:1.4.3}). Therefore
\[\Theta(E,x;v) = \delta(E_v,x)-(1-v)^2\delta(E,x)\propto(1-v)^2v/E^3\]
as $E\gg E_{\rm as}$, i.e. the function $\Theta(E,x;v)$ is small in absolute
value by comparison with $\delta(E,x)$. We assume that at all energies the term
with the factor $\Theta(E,x;v)$ can be neglected in the integrand of the right
side of Eq.~(\ref{e:1.6.5}) as a first approximation. Thus the equation for the
correction function in {\em second} approximation becomes
\be[\Li-R_2(E,x)]\delta^{(2)}(E,x)
              = R_0(E,x)\;,\;\;\;\;\;\delta^{(2)}(E,0) = 0\;,\label{e:1.6.7}\ee
where
\be R_l(E,x) = \int^1_0\Phi(v,E_v)\left\{\Omega(E,x;v)(1-v)^l -
                       [1+\omega(E,x)v]\right\}\eta(v,E)dv\;.\label{e:1.6.8}\ee
Solving Eq.~(\ref{e:1.6.7}) yields
\[\delta^{(2)}(E,x) = \int^x_0
     \exp\left[\int^x_{x'}R_2(\Ei(E,x-\Y),\Y)d\Y\right]R_0(\Ei(E,x-x'),x')dx'\]
\be\equiv\int^{\Ei(E,x)}_E\exp\left[\int^{\Ei(E,x)}_{E'}
          \frac{R_2(\W,x-\lambda_i(E,\W))}{\beta_i(\W)}d\W\right]
          \frac{R_0(E',x-\lambda_i(E,E'))}{\beta_i(E')}dE'\;,\label{e:1.6.9}\ee
where $\Ei(E,x)$ is the only root of the equation $\lambda_i(\Ei,E) = x$, and
\[\lambda_i(E_1,E_2) = \int^{E_1}_{E_2}\frac{dE}{\beta_i(E)}\]
is the ionization range of a muon with initial energy $E_1$ and final energy
$E_2$ (hence $\Ei(E,x)-E$ is simply the energy lost due to ionization).

Let us consider one evident consequence of Eq.~(\ref{e:1.6.9}). Clearly
$R_2(E,x) < R_0(E,x)$ for all values of the arguments. Substituting this
inequality into Eq.~(\ref{e:1.6.9}), and integrating over $E'$ then gives
\be\exp[K_2(E,x)]\leq
                    1+\delta^{(2)}(E,x)\leq\exp[K_0(E,x)]\;,\label{e:1.6.10}\ee
with\[K_l(E,x) = \int^x_0R_l(\Ei(E,x-x'),x')dx'\;.\]
The exponential factors in (\ref{e:1.6.10}) can be treated as the lower and
upper limits for the correction to the ``survival probability''
$\Prob$ so long as \[\int^x_0[q(\Ef(E,x-x'))-R_0(\Ei(E,x-x'),x')]dx' > 0\;.\]

In order to build an equation for calculation of the correction function in the
$l$-th approximation we note that the asymptotic behavior of the remainder
$\delta(E,x)-\delta^{(2)}(E,x)$ is $c_3(x)/E^3$ with an $E$-independent
function $c_3(x)$, as it can be easily verified using Eqs.~(\ref{e:1.6.5}) and
(\ref{e:1.6.7}). Therefore, in the next approximation we may put approximately
\[\delta(E_v,x)-\delta^{(2)}(E_v,x)\simeq
                                    (1-v)^3[\delta(E,x)-\delta^{(2)}(E,x)]\;.\]
Repeating the consideration we find by induction that $\delta(E,x) -
\delta^{(l)}(E,x)\rightarrow c_l(x)/E^l$ as $E\gg E_{\rm as}$, and thus we put
\be\delta(E_v,x)-\delta^{(l)}(E_v,x)\simeq
               (1-v)^{l+1}[\delta(E,x)-\delta^{(l)}(E,x)]\;.\label{e:1.6.11}\ee
Let us define
\be\Theta_l(E,x) = \delta^{(l)}(E,x)-\delta^{(l-1)}(E,x)\;,\;\;\;
                                                  l\geq 2\;,\label{e:1.6.12}\ee
with $\delta^{(1)}(E,x)\equiv 0$ by definition. From Eq.~(\ref{e:1.6.5}), using
Eq.~(\ref{e:1.6.12}) and Eq.~(\ref{e:1.6.11}) we obtain the recursion chain of
equations for the functions $\Theta_l(E,x)\,$:
\be [\Li-R_l(E,x)]\Theta_l(E,x)
                         = \Re_{l-1}(E,x)\;,\;\;\;l\geq 3\;,\label{e:1.6.13}\ee
where
\be\Re_l(E,x)=\int^1_0\Phi(v,E_v)\Omega(E,x;v)[\Theta_l(E_v,x)-
                         (1-v)^l\Theta_l(E,x)]\eta(v,E)dv\;.\label{e:1.6.14}\ee
The solution to Eq.~(\ref{e:1.6.13}) is given by
\[\Theta_l(E,x) = \int^x_0\exp\left[\int^x_{x'}R_l(\Ei(E,x-\Y),\Y)d\Y\right]
                                                 \Re_{l-1}(\Ei(E,x-x'),x')dx'\]
\be\equiv\int^{\Ei(E,x)}_E\exp\left[\int^{\Ei(E,x)}_{E'}
   \frac{ R_l     (\W,x-\lambda_i(E,\W))}{\beta_i(\W)}d\W\right]
   \frac{\Re_{l-1}(E',x-\lambda_i(E,E'))}{\beta_i(E')}dE'\;.\label{e:1.6.15}\ee

To verify the convergency of the iteration procedure consider firstly the
behavior of the function $R_l(E,x)$ at $l\gg 1$. Due to the factor $(1-v)^l$
in the first term of the integrand of Eq.~(\ref{e:1.6.8}) and the properties of
the macroscopic cross sections (see Sec.~\ref{ss:CS}), only the region of small
values of $v$ is important in this case. So at $l\gg 1$ the function
$\Omega(E,x;v)$ can be estimated as
\[\Omega(E,x;v) \simeq \Omega(E,x;0)+
                v\left[\frac{\partial \Omega(E,x;v)}{\partial v}\right]_{v=0}\]
or, considering the definitions (\ref{e:1.6.3}), (\ref{e:1.6.4}), and
(\ref{e:1.6.6}),
\be\Omega(E,x;v)\simeq 1+\omega(E,x)v \sim 1\;.\label{e:1.6.16}\ee Thus
\[R_l(E,x)\rightarrow
                   -\int^1_0\Phi(v,E_v)[1+\omega(E,x)v][1-(1-v)^l]\eta(v,E)dv\]
at $l\gg 1$ and, therefore, $R_l(E,x) < 0$ and $|R_l(E,x)|$ increases
indefinitely with $l$. Clearly the exponential factor
\[\exp\left[\int^x_{x'}R_l(\Ei(E,x-\Y),\Y)d\Y\right]\]
in the integrand of Eq.~(\ref{e:1.6.15}) diminishes fast with increasing $l$.

On the other hand, in view of the fact that $\Theta_l(E,x)\propto E^{-l}$ at
energies high enough, we can write
$\Theta_l(E_v,x)-(1-v)^l\Theta_l(E,x) = v(1-v)^l F_l(E,x;v)$,
where $F_l(E,x;v)$ is a function which can be estimated at $v\ll 1$ by
\[F_l(E,x;0) = E\frac{\partial\Theta_l(E,x)}{\partial E}+l\Theta_l(E,x)
                                                        \propto E^{-(l+1)}\;.\]
Hence, using Eq.~(\ref{e:1.6.16}), the integral (\ref{e:1.6.14}) can be
estimated at $l\gg 1$ as
\[\Re_l(E,x) \simeq
          F_l(E,x;0)\int^1_0\Phi(v,E_v)[1+\omega(E,x)v]\eta(v,E)(1-v)^lvdv\;.\]
Thus $\Re_l(E,x)$ and $\Re_l(\Ei(E,x-x'),x')$ are positive and decrease when
$l$ increases, if the function $F_l(E,x;0)$ is bounded in magnitude as
$l\rightarrow\infty$ or even if $|F_l(E,x;0)|$ increases with $l$ not too fast.
This can be verified by induction.

The foregoing proves that $\Theta_l(E,x)\rightarrow 0$ as $l\rightarrow\infty$
for any depths at least for high enough energies. Due to the correct asymptotic
behavior of the functions $\Theta_l(E,x)$ at all values of $l$ this indicates
that the iteration procedure converges, that is
\[\delta(E,x) = \lim_{l\rightarrow\infty}\delta^{(l)}(E,x)\;.\]
A more cumbersome analysis and numerical verifications demonstrate that this
statements is true {\em at all energies} under quite general assumptions on the
energy dependence of the rate of the continuous energy loss, the macroscopic
cross sections, and the initial spectrum; specifically if the functions
$\beta_i(E)$ and $\Phi(v,E)$ increase monotonically and sufficiently slowly,
while $D_0(E)$ decreases with energy so that $g(E)$ is a slightly varying
function of energy. As it follows from numerical estimations, the convergency
rate is very good, and usually only a few iterations are needed to reach an
accuracy of the order of 1\% at depths up to $\sim$ 20 km of water equivalent
for muon energies above $\sim$ 1 GeV.

By way of illustration let us again direct our attention to the SPS model and
consider the second approximation correction function $\delta^{(2)}_{\s}(E,x)$.
Under condition (\ref{e:1.4.6}) we can write the exact expression for the
correction function, using definition (\ref{e:1.6.1}) and Eqs.~(\ref{e:1.4.3})
and (\ref{e:1.5.5}):
\be\delta_{\s}(E,x) =
    \sum_{n=2}^{\infty}\frac{(\gamma+1)_n}{n!}\,\frac{f_n(x)-[f_1(x)]^n}
                    {[1+\xi f_1(x)]^{\gamma+1}}\,(-\xi)^n\;.\label{e:1.6.17}\ee
Therefore at $\zeta\leq 1$ the leading asymptotic term has the form
\be c_{\gamma}\epsilon\left[a_0+a_1e^{-\T}+a_1e^{-2\T}-
             (a_0+a_1+a_2)e^{-(2-\epsilon)\T}\right]\xi^2\;,\label{e:1.6.18}\ee
where $c_{\gamma} = (\gamma+1)(\gamma+2)/2$, $a_0 = 1/(2-\epsilon)$,
$a_1 = -2/(1-\epsilon)$, $a_2 = -1/\epsilon$, and $\epsilon = \Dgs/\Dgf$. The
expression for $R_l(E,x)$ in the SPS is given by
\be R_l(E,x) = R_l^{\gamma}(Z(E,x))\label{e:1.6.19}\ee with
\[R_l^{\gamma}(z) = \int^1_0\Phi(v)\left\{\frac{(1-v)^j}{(1-zv)^{\gamma+1}}
                                 -[1+(\gamma+1)zv]\right\}(1-v)^{\gamma}dv\;.\]
In particular, at $z \ll 1$ we have
\be R_0^{\gamma}(z)\simeq c_{\gamma}\Dgs z^2\;,\;\;\;\;
                   R_2^{\gamma}(z)\simeq-(2-\epsilon)\Dgf\;.\label{e:1.6.20}\ee
One can easily show, using Eqs.~(\ref{e:1.6.19}) and (\ref{e:1.6.20}) that the
asymptotic behavior of the exact correction function (\ref{e:1.6.18}) is
reproduced, as was to be expected, by the correction function in the second
approximation (see Eq.~(\ref{e:1.6.9})). It is important that the approximate
correction $\delta^{(2)}_{\s}(E,x)$ is definite and bounded at all values of
$E$ and $x$, in contrast with the exact expression (\ref{e:1.6.17}) given by a
series which converges only under condition (\ref{e:1.4.6}). This can be seen
from the constraint (\ref{e:1.6.10}). Indeed, a little manipulation, and taking
into account that
\[\partial Z(E,x)/\partial E\leq 0\;\;{\rm and}
                                         \;\;\partial Z(E,x)/\partial x\geq 0\]
yields
\bad K_0(E,x)& = &\:\sum^{\infty}_{n=2}\frac
{(\gamma+1)_n}{n!}\Dgn\int^x_0Z^n(E+a x',x-x')dx'\nonumber\\
& \leq &\:xR_0^{\gamma}(Z(E,x))Z(E+a x,x)/Z(E,x)\nonumber\\
& \leq &\:xR_0^{\gamma}\left(\frac{\xi}{1+\xi}\right)\frac{1+\xi}{1+\xi+\zeta}
                                          \equiv K_0^{\max}(E,x)\;.\nonumber\ea
Hence \[0\leq\delta^{(2)}_{\s}(E,x)\leq\exp[K_0^{\max}(E,x)]-1\;.\] and,
therefore, the survival probability calculated in the second approximation,
does not exceed the factor $\exp\{-[q_{\gamma}x-K_0^{\max}(E,x)]\}$. It is
obvious that the function $K_0^{\max}(E,x)$ and, therefore, the correction
$\delta^{(2)}_{\s}(E,x)$ are both bounded at all finite values of $E$ and $x$.
A significant consequence resides in the fact that $\delta^{(2)}_{\s}\ll 1$ at
$\xi\ll 1$ for any $x$ since
$K_0^{\max}\simeq c_{\gamma}\epsilon\xi\zeta/(1+\zeta)\ll 1$. In other words,
the first approximation solution (\ref{e:1.5.5}) is correct under the
condition $\xi\ll 1$ at all depths.

Analogous statements can be also proved in the general case: the first
approximation solution (\ref{e:1.5.3}) is practically exact {\em at all depths}
when $E\gg\beta_i(E)/\Delta_1(E)$.

\newpage\section{\bf SUMMARY AND OUTLOOK}\label{s:SO}

The method described enables us to calculate with a controlled precision the
differential energy spectra of CR muons after propagation through thick layers
of matter. It is appropriate for any depths when muon energies are high enough
and provides a way for including the real (non-power-law) initial muon
spectrum and the energy variation of the continuous and discrete muon energy
losses, with only the formal (but rather natural) requirement for the
asymptotic behavior of the initial spectrum and the cross sections. A computer
implementation of the method is fully straightforward and the required CPU time
is small. This enables to use the method in on-line processing for
underground/underwater experiments. It is important that the useful notion of
the minimal muon energy at the boundary $\Ec(E,x)$, which has a physical sense
when the range straggling is negligible (i.e. for small depths), has an analog
($\Ef(E,x)$) in the case of {\em arbitrary large fluctuations} (i.e. for
arbitrary depths) when the first approximation solution $\Df$ is available
(high muon energies, specifically $E >$ a few TeV).

We intend in the near future to give a detailed numerical illustration of the
convergency of the iteration procedure besides calculating results on the CR
muon energy spectra (differential and integral) and DIR at large depths
for different types of rocks and water with varying charm production models
{\em etc}. In recent years a rather representative array of data on DIR in rock
and (to a lesser extent) in water has been accumulated by many experiments. One
should systematize all these data and compare them against theoretical
predictions.

We also look forward to analyze future underground and underwater experiments
to possibly throw light on the prompt muon problem and to derive information
about super-high-energy muon interactions with nuclei, primarily about the
rather poorly studied photonuclear interaction.

\newpage\bc{\bf ACKNOWLEDGMENTS}\ec

We would like to thank Professor Akeo Misaki for useful discussions. One of
the authors (V. N.) is very grateful to Professor Bianca Monteleoni for many
helpful discussions, reading the manuscript, and critical remarks. V. N. also
thanks Professor Piero Spillantini for his continual encouragement and Mrs.
Silvia Cappelli for her skilled assistance in making the
typescript\vspace{10 mm}.

\bc{\bf APPENDIX\2:}\ec\bc{\bf\1\half van Ginneken's parameterization of\1
                          the v-distributions \\ for radiative \2 processes}\ec

We give here a parameterization of the normalized cross sections $f_k(v,E) =
\sigma_k^{-1}d\sigma_k/dv\;$ (where $d\sigma_k/dv\equiv d\sigma_k^{Z,A}/dv)$
for pair production and bremsstrahlung as a function of the fractional energy
loss $v$ suggested in Ref.~\cite{vG86}. The formulas presented below are valid
at muon energies from $\sim 100$ GeV up to 30 TeV and enable one to estimate
the comparative probabilities of ``soft'' ($v\ll 1$) and ``hard'' ($v\sim
v_{\max}^k\sim 1$) losses in the radiative processes.

 \bc\bb{\sf Direct pair production off both nuclear and electron targets}\eb\ec
\bA{lclrcl} f_p^{(n,e)}(v,E)
\pr const\,,\;\;&  5m_e/E <&v&<  25m_e/E\,,\\
\pr v^{-1}\,,   & 25m_e/E <&v&< v_1\;\;({\rm if}\;\;E>25m_e/v_1)\,,\\
\pr v^{-2}\,,   &     v_1 <&v&< v_2\;\;({\rm if}\;\;E>25m_e/v_2)\,,\\
\pr v^{-3}\,,   &     v_2 <&v&< 1\,,                                        \eA
where $\,v_1 = 0.002\,$ and $\,v_2 = 0.02\,$. For very small $v$ (up to the
kinematic limit $v^p_{\min}(E) = 4m_e/E$) $\;d\sigma_p(v,E)/dv\propto
\sigma_0(vE)\ln(1/v)v^{-1}$, where $\sigma_0(\nu)$ is the total cross section
for pair production by a photon of energy $\nu$ (Kel'ner's approximation). Thus
$d\sigma_p/dv$ follows roughly a $1/v$-dependence in the region $\nu\gg 1$ MeV,
where $\sigma_0(\nu)$ is practically constant, $\sigma_0\approx
\sigma_0(\infty)$. Below $\nu\approx 5m_e\,,\;\,\sigma_0(\nu)$ remains small
($< 0.05\sigma_0(\infty)$), and it increases roughly linearly until
$\nu\approx 25m_e$, where $\sigma_0(\nu)\approx 0.5\sigma_0(\infty)$.

    \bc\bb{\sf Bremsstrahlung contribution off a nuclear target}\eb\ec
\bA{lclrcl} f_b^{(n)}(v,E)
\pr v^{-1}\,,                  & v^b_{\min}(E) <&v&< 0.03\,,\\
\pr v^{C_n(E)}\,,              & 0.03 <&v&< v_b^n(E)\,,\\
\pr (1-v)^{C'_n(E)}\,,\;\;\;   & v_b^n(E) <&v&< 1\,,                        \eA
where $\,v^b_{\min}(E) = 0.001/E\,,\;\;v_b^n(E) = (1+4.5/\sqrt{E})^{-1}\,,\;\;
C_n(E) = 1.39-0.024\ln E\,$, and $\,C'_n(E) = 1.32-0.12\ln E\,$. Only for
values of $v$ above 0.995 the parameterization is not very reliable, but it is
not important in practice.

     \bc\bb{\sf Bremsstrahlung contribution off atomic electrons}\eb\ec
\bA{lclrcl} f_b^{(e)}(v,E)
\pr v^{-0.95}\,,                    & v^b_{\min}(E) <&v&< 0.05\,,\\
\pr v^{C_e(E)}\,,                   & 0.05 <&v&< v_b^e(E)\,,\\
\pr (v^b_{\max}(E)-v)^{1/2}\,,\;\;\;& v_b^e(E) <&v&< v^b_{\max}(E)\,,       \eA
where $\,v^b_{\max}(E) = (1+10.92/E)^{-1}\,,\;\;v_b^e(E) =
v_b^n(E)v^b_{\max}(E)\,$, and $\,C_e(E) = 1.50-0.03\ln E\,$.

The proportionality factors which was omitted in the above parameterization can
be obtained by continuity and normalization. A slight $Z$-dependence in the
$v$-distributions has been ignored. The energy $E$ and the electron mass $m_e$
have been expressed in GeV.

\end{document}